\documentclass[12pt,a4paper]{article}
\usepackage{amsmath}
\usepackage[dvipdfmx]{graphicx}
\begin{document}

%\draft

\title{
Spectral function method for Hall conductivity of incoherent metals 
}

\author{
Osamu Narikiyo\\ 
{\it Department of Physics, 
Kyushu University, 
Fukuoka 819-0395, 
Japan}\\ 
{\it narikiyo@phys.kyushu-u.ac.jp}
}

\date{
(24 August 2016)
}

\maketitle
%----------------------------------------------------------------
\begin{abstract}
\noindent
Using a model spectral function of the electron 
the Hall conductivity 
in the normal metallic state of the Pr$_{2-x}$Ce$_x$CuO$_4$ (PCCO) superconductor 
is calculated neglecting the current vertex-correction. 
The result is qualitatively consistent with the experiment. 
Consequently the reason becomes clear 
why the Fermi-liquid theory fails to explain the anomaly of the Hall conductivity. 
The inconsistency of the fluctuation-exchange approximation also becomes clear. 

\vskip 15pt 


\vskip 15pt 


\end{abstract}
%----------------------------------------------------------------

\vskip 30pt 

\noindent
{\bf 1. Introduction}

\vskip 5pt

\noindent
Various anomalous properties 
in the normal metallic state of cuprate superconductors 
cannot be explained by the Fermi-liquid theory 
which is the standard theory for interacting electrons.
One of the most striking and puzzling anomalies 
is the strong temperature dependence of the Hall coefficient~\cite{Anderson1991}. 
In the Fermi-liquid theory for a single-band case 
the Hall coefficient is a measure for the carrier density of the metal 
and almost temperature-independent~\cite{KY1988}. 
Thus many attempts have been made to explain the Hall coefficient 
from non-Fermi-liquid viewpoints~\cite{Anderson1991,Anderson1997,LNW2006}. 
However, such attempts seem to be unsuccessful. 
On the other hand, 
it is claimed using the fluctuation-exchange (FLEX) approximation~\cite{Kontani2008} 
that the anomalous behavior of the coefficient (ABC) comes 
from the current vertex-correction (CVC). 
Since the Hall coefficient of the FLEX approximation 
is evaluated within the Fermi-liquid framework, 
such a claim contradicts the standard argument 
leading to almost temperature-independent Hall coefficient. 

Recently the transport anomalies have been discussed on the basis of 
hidden Fermi-liquid~\cite{Anderson2009,Anderson2011} or 
anisotropic marginal Fermi-liquid~\cite{KM2011,KHM2012} model. 
In these discussions the anisotropy in 
the temperature-dependence of the imaginary part of the electron self-energy 
in momentum space is crucial 
but the CVC is neglected. 
To explain the Hall coefficient 
the anisotropy of the real part is also important as discussed in the following. 
Anyway, the essential part of the transport coefficients 
is expressed in terms of the spectral function of the electron 
whose non-trivial part is characterized by the self-energy. 
For example, this is proved experimentally~\cite{ARPES2008} for a CDW pseudo-gap material 
where an anomalous temperature dependence of the Hall coefficient is explained 
using the spectral function 
measured by the angle-resolved photoemission spectroscopy (ARPES). 
We support this spectral function method. 

In this letter 
we demonstrate the effectiveness of the spectral function method 
for the Hall conductivity using a simple model of the spectral function. 
At the same time 
it becomes clear why the Fermi-liquid theory fails 
to explain the anomaly of the Hall conductivity. 
We also show the inconsistency of the FLEX approximation. 
Since the comparison of our result 
and that by the FLEX approximation~\cite{Drew2010} 
for the Pr$_{2-x}$Ce$_x$CuO$_4$ (PCCO) superconductor reveals the inconsistency, 
we choose this material as the target of our calculation. 

\vskip 15pt 

\noindent
{\bf 2. Conductivity formula}

\vskip 5pt

\noindent
We start from the full Green function for electrons 
\begin{equation}
G^R({\bf p},\varepsilon) 
= {1 \over \varepsilon - E({\bf p},\varepsilon) + i/2\tau({\bf p},\varepsilon)}, 
\end{equation}
or 
\begin{equation}
G^A({\bf p},\varepsilon) 
= {1 \over \varepsilon - E({\bf p},\varepsilon) - i/2\tau({\bf p},\varepsilon)}, 
\end{equation}
where the real-part of the self-energy is renormalized into 
the dispersion $E({\bf p},\varepsilon)$ and 
the imaginary-part into the life-time $\tau({\bf p},\varepsilon)$. 
Here we have made no assumption on the Green function. 
Namely, the following discussion is free from the nature of the quasi-particle, 
Fermi liquid or non-Fermi liquid. 

The spectral function of the electron is given as 
\begin{equation}
\rho_{\bf p}(\varepsilon) 
= {1 \over \pi}
{1/2\tau({\bf p},\varepsilon) \over \big[ \varepsilon - E({\bf p},\varepsilon) \big]^2 
+ \big[1/2\tau({\bf p},\varepsilon) \big]^2 }. 
\end{equation}
The DC conductivities are essentially determined by this function at $\varepsilon = 0$ 
as discussed in the following. 
 
The integrand of the momentum-integration of the DC conductivity $\sigma_{xx}$ 
in the absence of the magnetic field is proportional to $v_x \cdot v_x$, 
where $v_x$ is the $x$-component of the group velocity of the renormalized quasi-particle, 
and the effect of the CVC 
by the residual interaction among quasi-particles 
is expressed as the renormalization of the life-time $\tau({\bf p},\varepsilon)$ 
into the transport life-time. 
The proof of this is given in~\cite{YY1986} for spherically symmetric case 
and in~\cite{MF1998} for anisotropic case. 
The same is true for the DC Hall conductivity $\sigma_{xy}$ 
linear in the magnetic field $H$. 
The proof is given in~\cite{KF2003} for spherically symmetric case 
and in~\cite{Narikiyo2014} for anisotropic case. 
The FLEX approximation~\cite{Kontani2008} cannot meet this exact result, 
since it violates the harmony~\cite{YY1986,MF1998} 
between the imaginary part of the self-energy and the CVC. 
Such a harmony is a consequence of the local electron-number conservation~\cite{MF1997}. 
However, the Baym-Kadanoff scheme including the FLEX approximation 
only assures the global conservation~\cite{MT2011}. 

Although the renormalization in the transport life-time 
has a significant effect on the conductivities 
by removing forward scattering contributions 
irrelevant to the change in the velocity of electrons 
in the case where some long-wavelength scattering 
(see \cite{Ziman1960} for phonons 
and \cite{Moriya1985} for ferromagnetic spin-fluctuations) 
is dominant, 
we can neglect~\cite{HR1995,SP1997} such a renormalization for our target material, 
since the scattering across the Fermi surface is dominant in our case. 

Our numerical calculation is done for 2D square lattice 
with the lattice constant $a \equiv 1$. 
By the symmetry we only need the information 
for the quarter of the Brillouin zone: 
$0 \leq p_x \leq \pi$ and $0 \leq p_y \leq \pi$. 
In the following 
the summation over ${\bf p}$ is restricted within this quarter. 

Neglecting the CVC 
the DC conductivity $\sigma_{xx}$ in the absence of the magnetic field 
is given by~\cite{FEW1969} 
\begin{equation}
\sigma_{xx} 
= {e^2 \over \pi} \cdot {4 \over R(T)}, 
\label{sigma_xx} 
\end{equation}
with 
\begin{equation}
{1 \over R(T)} 
\equiv \sum_{\bf p} v_x^2 G^R G^A 
= \sum_{\bf p} v_x^2 {1 \over E^2 + (1/2\tau)^2}, 
\label{R(T)} 
\end{equation}
where 
$G^R=G^R({\bf p},0)$, 
$G^A=G^A({\bf p},0)$, 
$E=E({\bf p},0)$, 
$\tau=\tau({\bf p},0)$ and 
$v_x=\partial E / \partial p_x$. 
Here we have assumed the Fermi degeneracy for simplicity. 

In the same manner 
the DC Hall conductivity $\sigma_{xy}$ 
proportional to the weak magnetic field $H$ 
is given by \cite{FEW1969} 
\begin{equation}
\sigma_{xy} / H 
= {|e|^3 \over \pi} \cdot 4 S(T), 
\label{sigma_xy} 
\end{equation}
with 
\begin{align}
S(T) 
& \equiv \sum_{\bf p} 
v_x \Big( v_x {\partial v_y \over \partial p_y} 
        - v_y {\partial v_x \over \partial p_y} \Big) 
G^R G^A { G^R - G^A \over 2 i } 
\nonumber \\ 
& = - \sum_{\bf p} 
v_x \Big( v_x {\partial v_y \over \partial p_y} 
        - v_y {\partial v_x \over \partial p_y} \Big) 
{1/2\tau \over \big[ E^2 + (1/2\tau)^2 \big]^2}, 
\label{S(T)} 
\end{align}
where $v_y=\partial E / \partial p_y$. 
Thus, if we have the data of 
the dispersion $E$ and the life-time $\tau$ at $\epsilon=0$, 
we can get the DC conductivities from these formulae. 
The necessary data are obtained from the spectral function. 

If the ARPES supplies such data, 
we can immediately calculate the conductivities. 
However, the purpose of this letter is to demonstrate the effectiveness 
of the spectral-function method so that 
we phenomenologically use a simple model for the spectral function for simplicity. 
To derive the model spectral function from a microscopic model 
is another task to do but not discussed in this letter. 
Although we introduce the model spectral function under the assumption 
that the dominant scattering is due to the anti-ferromagnetic spin-fluctuation, 
some qualitatively similar spectral function can be obtained 
from the other frameworks~\cite{Anderson2009,Anderson2011,KM2011,KHM2012}. 

\vskip 15pt 

\noindent
{\bf 3. Model spectral function}

\vskip 5pt

\noindent
Here we use 
the dispersion for renormalized quasi-particles 
\begin{equation}
E = -2t(\cos p_x + \cos p_y) + 4t'\cos p_x \cdot \cos p_y 
    -2t''(\cos 2p_x + \cos 2p_y) - \mu, 
\label{E(p)} 
\end{equation}
obtained by the band calculation~\cite{NKS2011}.  
For PCCO we adopt $t=-0.438$eV, $t'=0.156$eV and $t''=0.098$eV~\cite{NKS2011} 
in the numerical calculations in this letter. 

The Fermi surface for this dispersion is shown in Fig.~1. 
The factor $f({\bf p}) \equiv 
 v_x ( v_x \partial v_y / \partial p_y 
     - v_y \partial v_x / \partial p_y ) $ 
appearing in $\sigma_{xy}$ is shown in Fig.~2. 

For the life-time we adopt the model 
\begin{equation}
{1 \over 2\tau} = 
w({\bf p})\cdot g_1 \cdot T 
+ \big[ 1 - w({\bf p}) \big]\cdot g_2 \cdot T^2, 
\label{tau} 
\end{equation}
similar to the multi-patch model~\cite{PSK2001} 
where the Brillouin zone is divided into hot and cold patches. 
We expect that the life-time is determined 
by the coupling to the anti-ferromagnetic spin-fluctuation. 
In the cold patches 
the relevant spin-fluctuation spectrum is broad 
and its temperature-dependence is weak 
so that $1/\tau \sim T^2$~\cite{HR1995,SP1997,NM1998}. 
On the other hand, in the hot patches 
it is peaked around the nesting-vector 
and its integrated weight depends on the temperature 
so that $1/\tau \sim T$~\cite{HR1995,SP1997,NM1998}. 
For simplicity 
we did not implement the gradual change 
between hot and cold patches employed in~\cite{PSK2001}. 
Namely $w({\bf p})$ is a step function in our case 
where $w({\bf p})=1$ 
if $({\bf p}-{\bf p}_1)^2 < r^2$ or $({\bf p}-{\bf p}_2)^2 < r^2$ 
with ${\bf p}_1 \equiv (\pi,0)$ and ${\bf p}_2 \equiv (0,\pi)$ 
and $w({\bf p})=0$ otherwise. 
The parameter $r^2$ is chosen to be $1.07$. 

We set $g_2 = 100/$eV in accordance with~\cite{KHM2012}. 
Here the temperature $T$ is measured in eV. 
On the other hand, our choice, $g_1 = 0.1$, is smaller 
than~\cite{KHM2012} and~\cite{PSK2001}. 
Such a smallness is explained by the nested spin fluctuation~\cite{NM1998}. 

\vskip 15pt 

\noindent
{\bf 4. Numerical results}

\vskip 5pt

\noindent
Before performing the numerical calculation 
we can make a rosy prediction 
that the non-monotonic temperature dependence of $\sigma_{xy}$, 
which we want to derive, is obtained, 
if the temperature dependences of $1/\tau$ are different 
between the regions with positive $f({\bf p})$ 
and the regions with negative $f({\bf p})$ 
in the Brillouin zone. 
After the numerical calculation 
we confirm the prediction as shown in Fig.~3. 
The result for $\mu=0$ actually shows 
the non-monotonic temperature dependence 
qualitatively similar to that observed in the experiment~\cite{Drew2010}. 
More direct comparison between the experiment~\cite{Drew2010} 
and our numerical calculation can be done 
for $\sigma_{xy}/\sigma_{xx}$ shown in Fig.~4 
which qualitatively explains the experiment. 

\vskip 15pt 

\noindent
{\bf 5. Remarks}

\vskip 5pt

\noindent
Our simple formula based on the spectral function of the electron 
qualitatively explains the non-monotonic temperature-dependence 
of the DC Hall conductivity in the normal state of PCCO superconductors. 

As shown in~\cite{FEW1969} 
this formula leads to the constant Hall coefficient 
if the spectral function $\rho_{\bf p}(\varepsilon)$ is delta-function-like 
and isotropic in momentum space. 
The Fermi-liquid theory for the Hall coefficient~\cite{KY1988} also assumes 
the delta-function-like spectral function 
so that it also leads to almost temperature-independent Hall coefficient. 
While the Fermi-liquid theory exploits only the information on the Fermi surface, 
our scheme exploits the entire Brillouin zone. 
Thus the Fermi-liquid theory fails to explain the ABC. 
Although some anisotropy leads to 
weak temperature dependence of the Hall coefficient~\cite{Narikiyo2000}, 
it is too weak to explain the anomalous temperature dependence 
observed in experiments. 

The above-mentioned delta-function-like spectral function 
is justified only for weakly correlated systems 
in infinite momentum space at extremely low temperatures. 
Our calculation in the main text 
is performed without such a delta-function assumption. 
We have to use a broad spectral function 
in a finite momentum space (the 1st Brillouin zone) 
in the case of strongly correlated system at the room temperature. 
Thus the understanding of the ABC is out of the scope of the Fermi-liquid theory. 
Our expecting spectral function is schematically shown in Fig.~5. 

The normal state of cuprate superconductors is 
an incoherent metal~\cite{mFL,Imada,inc1,inc2}. 
The spectral function method discussed in this letter is an attempt 
to describe the transport properties of incoherent metals 
where we have to consider the entire Brillouin zone not only the Fermi surface. 
The other approaches~\cite{Hartnoll,Sachdev}  
to incoherent metals are also intensively discussed recently. 

Although our spectral function employed in this letter is a simple model, 
it captures the features of actual one observed by the ARPES experiments~\cite{NM1998}. 
Thus our result for the Hall angle in Fig.~4 
qualitatively explains the anomalous non-monotonic temperature dependence 
observed in experiments. 
Thus we can conclude 
that the ABC is explained without the CVC 
if we employ the correct spectral function. 

On the other hand, 
the Hall angle calculated 
without the CVC in the FLEX approximation, 
the inset of Fig.~2 in~\cite{Drew2010}, 
is totally different from the one observed by experiments. 
The failure of this calculation does not mean the necessity of the CVC. 
It only means 
that the spectral function obtained by the FLEX approximation is incorrect. 
The reason for the incorrectness is widely known~\cite{VT1997}. 
We have also discussed~\cite{Narikiyo2013} 
that the FLEX approximation is not applicable 
to the system with the Fermi degeneracy, 
since it violates the Pauli principle~\cite{VT1997} 
and the local electron-number conservation~\cite{MT2011}. 

Consequently, the spectral function is the key to explain the ABC. 
The FLEX approximation fails to explain the ABC, 
since the correct spectral function is not obtained by this approximation. 
The Fermi-liquid theory fails to explain the ABC, 
since the delta-function assumption is not justified 
for strongly correlated systems at the room temperature. 

The author is grateful to Kazumasa Miyake 
for illuminating discussions for two decades. 

\vskip 15pt 

%----------------------------------------------------------------

\newpage

%%%%%%%%%%%%%%%%%%%%%%%%%%%%%%%%%%%%%%%%%%
\begin{figure}
\centering
\includegraphics[width=12.0cm,height=12.0cm]{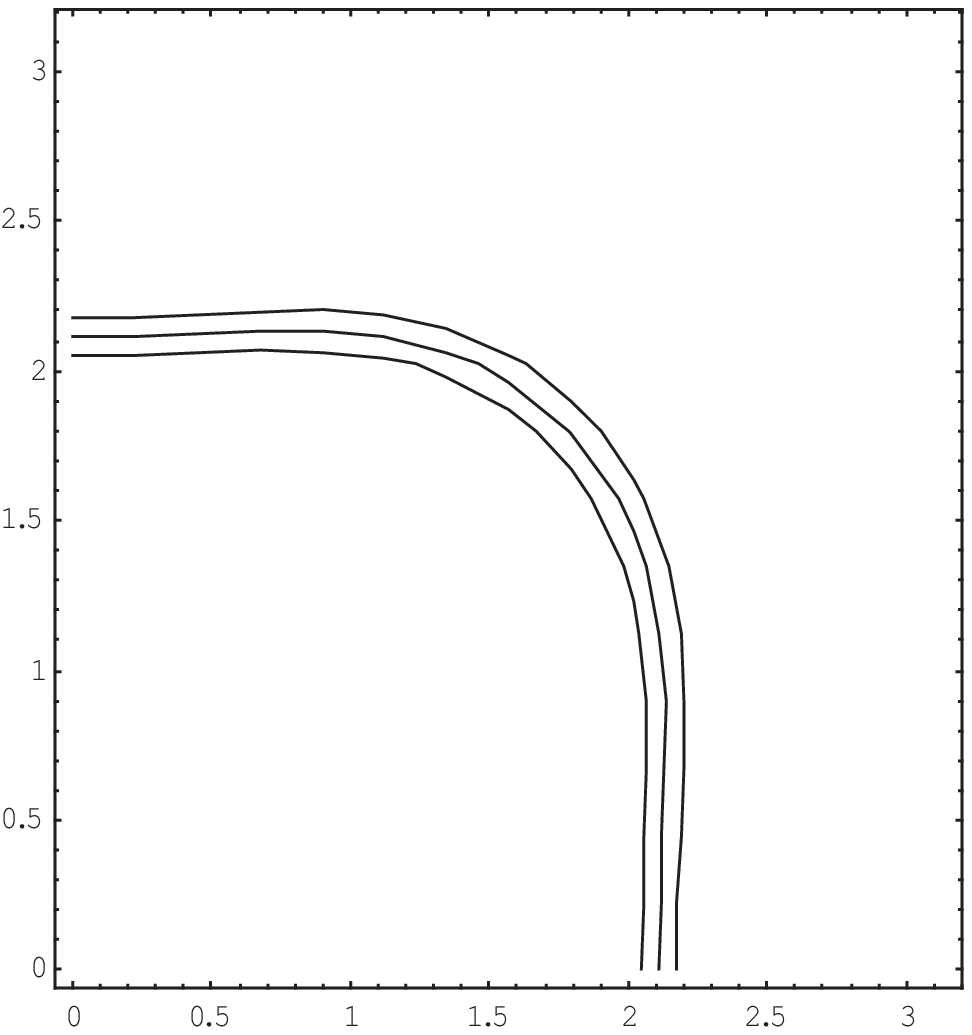}
\caption{Fermi surfaces for $\mu=-0.1$, $0$, $0.1$ 
with $t=-0.438$, $t'=0.156$, $t''=0.098$ 
where all the energies are represented in eV. 
Only the quarter of the Brillouin zone 
($0 \leq p_x \leq \pi$ and $0 \leq p_y \leq \pi$) 
is depicted. }
\label{fig:FS}
\end{figure}
%%%%%%%%%%%%%%%%%%%%%%%%%%%%%%%%%%%%%%%%%%

\newpage 

%%%%%%%%%%%%%%%%%%%%%%%%%%%%%%%%%%%%%%%%%%
\begin{figure}
\centering
\includegraphics[width=12.0cm,height=12.0cm]{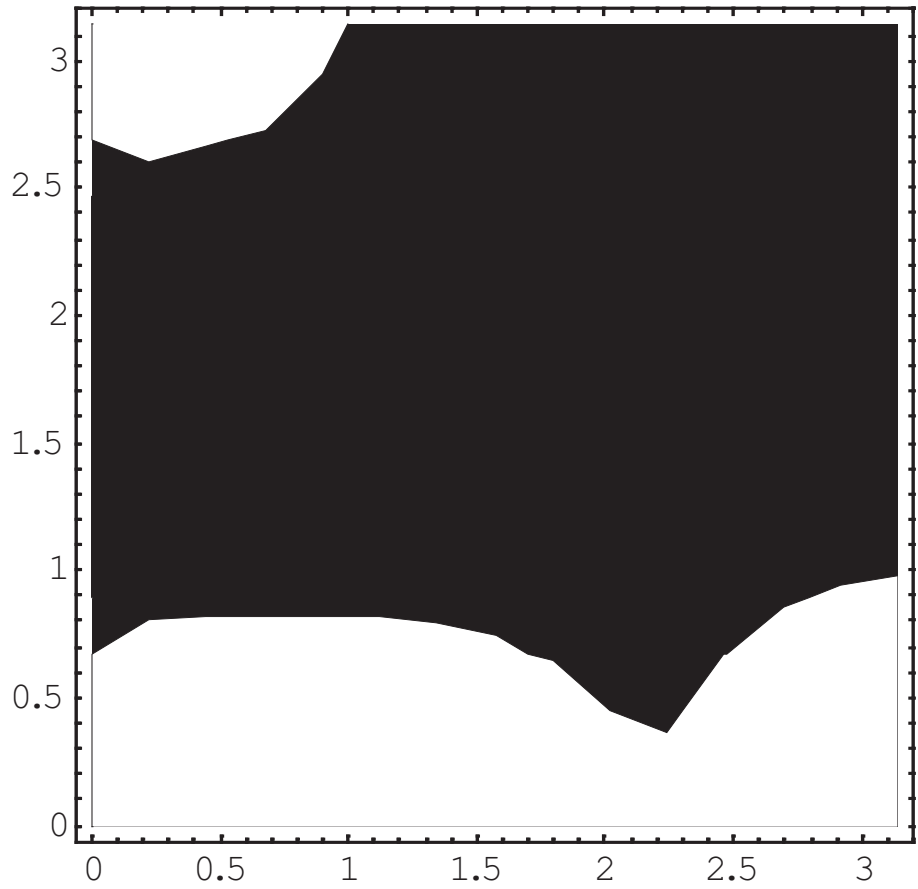}
\caption{The factor 
$f({\bf p}) = 
 v_x ( v_x \partial v_y / \partial p_y 
     - v_y \partial v_x / \partial p_y ) $ 
in the quarter of the Brillouin zone. 
In the white region $f({\bf p}) > 0$ 
and $f({\bf p}) < 0$ in the black region. }
\label{fig:curv}
\end{figure}
%%%%%%%%%%%%%%%%%%%%%%%%%%%%%%%%%%%%%%%%%%

\newpage 

%%%%%%%%%%%%%%%%%%%%%%%%%%%%%%%%%%%%%%%%%%
\begin{figure}
\centering
\includegraphics[width=16.0cm]{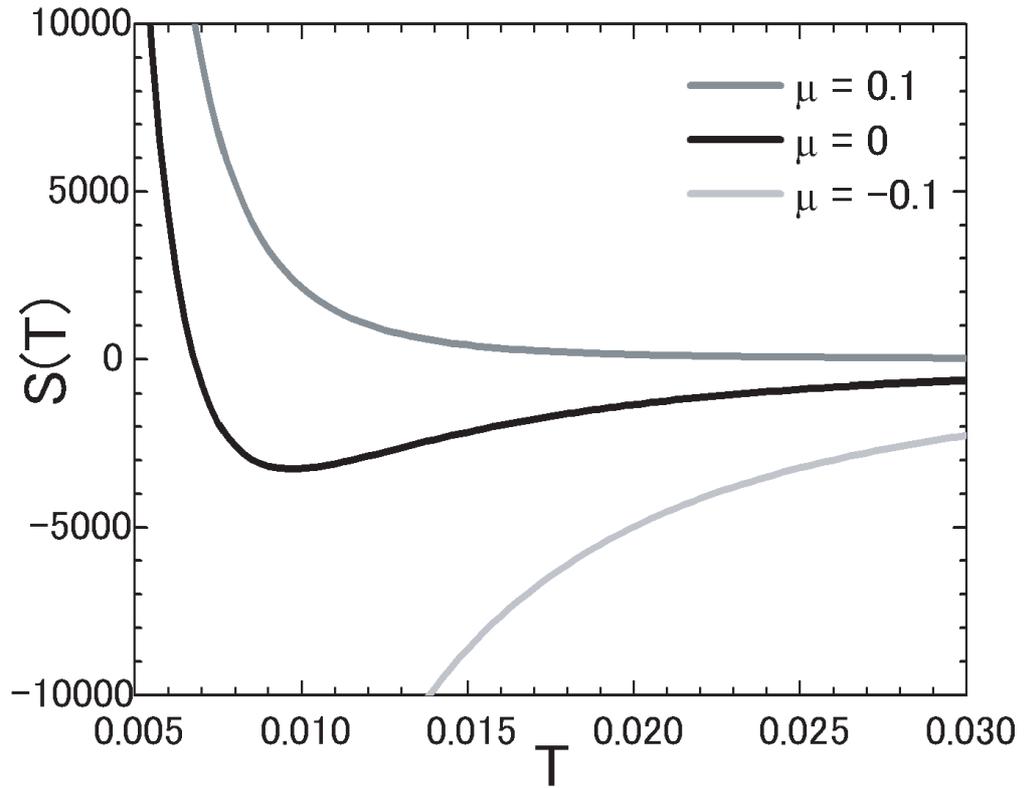}
\caption{Temperature dependencies of $S(T)$ 
for $\mu=-0.1$, $0$, $0.1$ with $r^2 = 1.07$ 
where $S(T) \propto \sigma_{xy}$. 
The temperature is represented in eV. 
The summation over ${\bf p}$ is carried out 
using 3000$\times$3000 mesh in the quarter of the Brillouin zone. }
\label{fig:sigma}
\end{figure}
%%%%%%%%%%%%%%%%%%%%%%%%%%%%%%%%%%%%%%%%%%

\newpage 

%%%%%%%%%%%%%%%%%%%%%%%%%%%%%%%%%%%%%%%%%%
\begin{figure}
\centering
\includegraphics[width=16.0cm]{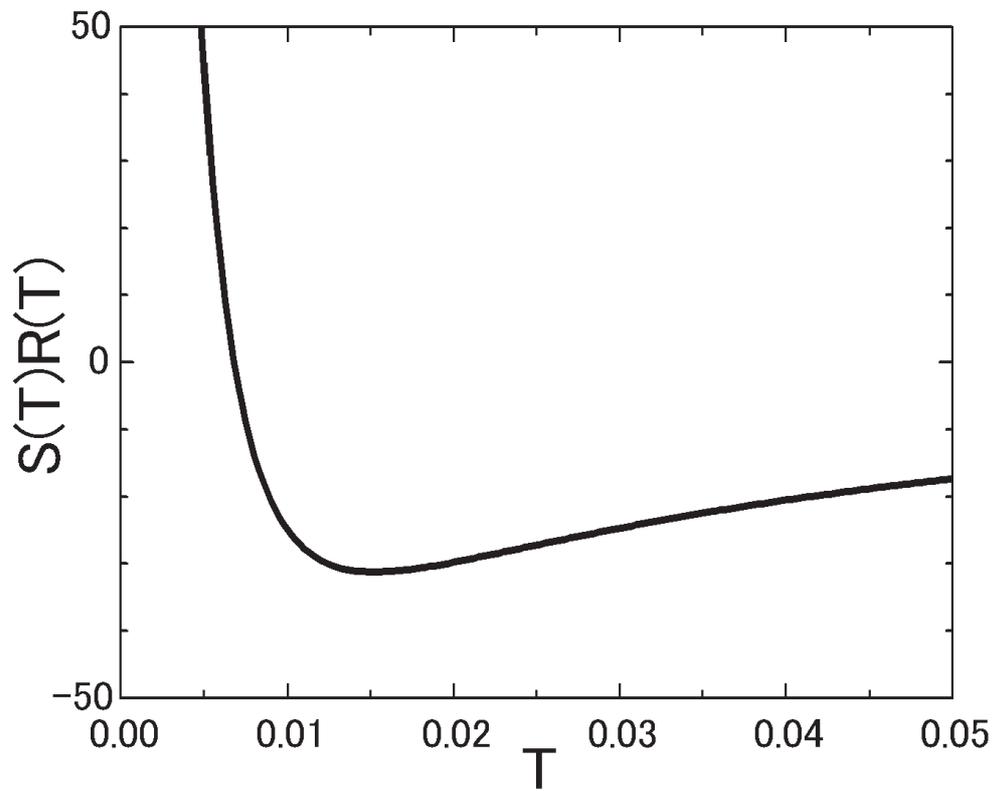}
\caption{The temperature dependence of $S(T)R(T)$ 
for $\mu=0$ with $r^2 = 1.07$ 
where $S(T)R(T) \propto \sigma_{xy}/\sigma_{xx}$. }
\label{fig:TH}
\end{figure}
%%%%%%%%%%%%%%%%%%%%%%%%%%%%%%%%%%%%%%%%%%

\newpage 

%%%%%%%%%%%%%%%%%%%%%%%%%%%%%%%%%%%%%%%%%%
\begin{figure}
\centering
\includegraphics[width=16.0cm]{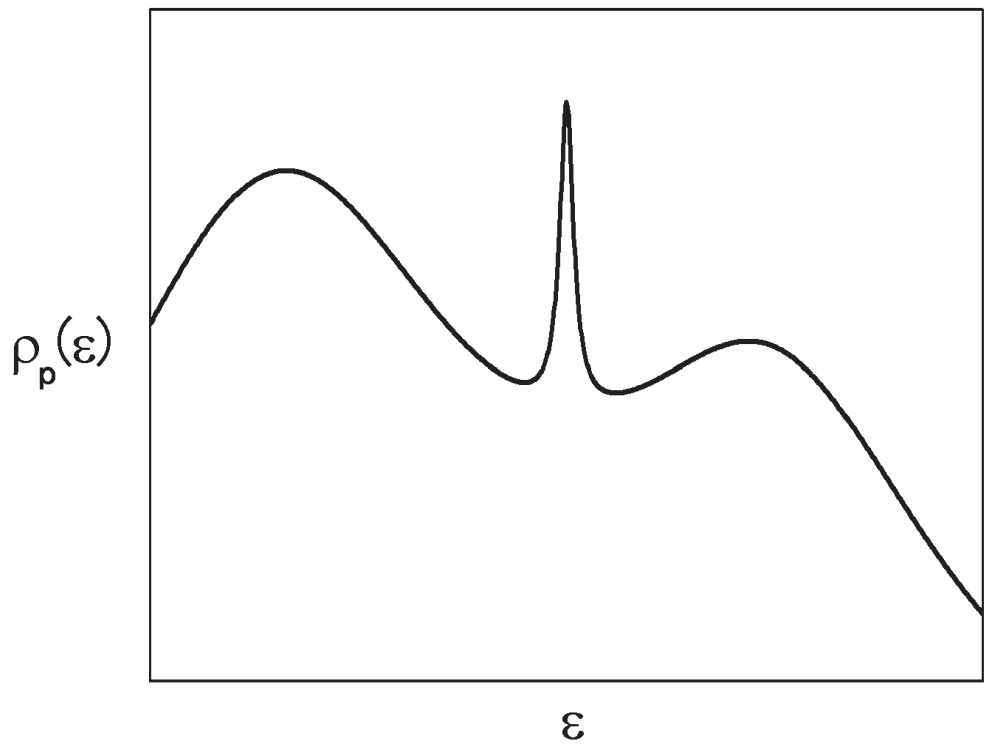}
\caption{
Energy dependence of spectral function. 
}
\label{fig:spec}
\end{figure}
%%%%%%%%%%%%%%%%%%%%%%%%%%%%%%%%%%%%%%%%%%

%----------------------------------------------------------------
\end{document}